\documentclass[fleqn,usenatbib]{mnras}

\usepackage{newtxtext,newtxmath}
\usepackage{ae,aecompl}
\usepackage[utf8]{inputenc}
\usepackage[version=3]{mhchem} 
\usepackage{siunitx}
\usepackage{graphicx}
\usepackage{xcolor}
\usepackage{enumerate}

\usepackage{color}

\title[H transfer of COM reactions]{Hydrogen transfer reactions of interstellar Complex Organic Molecules}

\author[\'{A}lvarez-Barcia et al.]{
S. \'{A}lvarez-Barcia,$^{1}$
P. Russ,$^{1}$
J. K\"{a}stner,$^{1}$
and T. Lamberts$^{1,2}$\thanks{E-mail: a.l.m.lamberts@lic.leidenuniv.nl}
\\
$^{1}${Institute for Theoretical Chemistry, University of Stuttgart, Pfaffenwaldring 55, 70569 Stuttgart, Germany}\\
$^{2}${Current Address: Leiden Institute of Chemistry, Gorlaeus Laboratories, Leiden University, P.O. Box 9502, 2300 RA Leiden, The Netherlands}
}

\date{Accepted XXX. Received YYY; in original form ZZZ}
\pubyear{2018}

\newcommand{\jpca}{J.~Phys.~Chem.~A}
\newcommand{\pccp}{Phys.~Chem.~Chem.~Phys.}%
 \newcommand{\aapl}{Astron.~Astrophys.~Lett.}
\begin{document}
\label{firstpage}
\pagerange{\pageref{firstpage}--\pageref{lastpage}}
\maketitle

\begin{abstract}
{Radical recombination has been proposed to lead to the formation of complex organic molecules (COMs) in CO-rich ices in the early stages of star formation. These COMs can then undergo hydrogen addition and abstraction reactions leading to a higher or lower degree of saturation. Here, we have studied  {14} hydrogen transfer reactions for the molecules glyoxal, glycoaldehyde, ethylene glycol, and methylformate  {and an additional three reactions where \ce{CH_nO} fragments are involved.} Over-the-barrier reactions are possible only if tunneling is invoked in the description at low temperature. Therefore the rate constants for the studied reactions are calculated using instanton theory that takes quantum effects into account inherently.  {{The reactions were characterized in the gas phase, but this is expected to yield meaningful results for CO-rich ices due to the minimal alteration of reaction landscapes by the CO molecules.}}

We found that rate constants should not be extrapolated based on the height of the barrier alone, since the shape of the barrier plays an increasingly larger role at decreasing temperature. It is neither possible to predict rate constants based only on considering the type of reaction, the specific reactants and functional groups play a crucial role. Within a single molecule, though, hydrogen abstraction from an aldehyde group seems to be always faster than hydrogen addition to the same carbon atom. Reactions that involve heavy-atom tunneling, \emph{e.g.}, breaking or forming a C--C or C--O bond, have rate constants that are much lower than those where H transfer is involved. 
}
\end{abstract}


\section{Introduction}
Thanks to the unprecedented sensitivity of ALMA the detection and quantification of interstellar complex organic molecules (COM) has become more and more within reach. A complex organic molecule in the context of astrochemistry is loosely defined as a molecule consisting of more than 6 H, C, O, and/or, N atoms. Typical gas-phase abundances of such molecules are only of the order of $< 10^{-8}$ with respect to \ce{H2} \citep{Jorgensen:2012,Halfen:2015,Taquet:2015,Lopez-Sepulcre:2017} with even lower abundances for deuterated species \citep{Belloche:2016}. These molecules are currently thought to find their origins in the CO-rich top layers of the grain ice mantle \citep{Boogert:2015} where the \ce{H + CO} reaction network has been shown to lead to the formation of the parent species formaldehyde, \ce{H2CO}, and methanol, \ce{CH3OH}  {\citep{Tielens:1982,Hiraoka:1998,Watanabe:2002,Fuchs:2009}}. Furthermore, besides hydrogen addition reactions, also hydrogen abstraction reactions can take place that decrease the number of H atoms on the CO backbone  {\citep{Nagaoka:2005,Nagaoka:2007}}. Although it has been suggested that formaldehyde and methanol may desorb from the grain surface and subsequently react in the gas phase to yield more complex species \citep{Bottinelli:2004, Balucani:2015,Taquet:2017}, a variety of COMs have been detected in cold interstellar regions \citep{Bacmann:2012,Oberg:2010,Vastel:2014}. This indicates that low-temperature surface chemistry can play an important role in the formation of larger species. 

In fact the \ce{H + CO} reaction network has evolved into a network where carbon-carbon bonds can be formed via radical-radical reactions between the `fundamental' radicals that are created as intermediates, \emph{i.e.}, HCO, \ce{CH2OH}, and \ce{CH3O}. Most of these reactions have been studied experimentally in various ways \citep{fed15,but15,chu16,fed17,chu17,but17} and they have also been proposed by and are included in a number of astrochemical model studies \citep{Garrod:2008,Woods:2012,Coutens:2018}. Similar conclusions are also supported by observational work for specific species \citep{Rivilla:2017,Li:2017}. Despite this significant amount of investigations, relatively little is known about the reaction rate constants at low temperature, while these are the crucial parameters needed to constrain modeling studies.

Here we focus on hydrogen addition and abstraction reactions of species with two carbon atoms and two oxygen atoms, \emph{i.e.}, methylformate, glyoxal, glycoaldehyde, and ethylene glycol (Section~\ref{sec:network-new}).  {Several other reactions are discussed as well where a C--C or C--O bond is formed via an over-the-barrier reaction between an \ce{CH_nO} radical and formaldehyde (Section~\ref{sec:FA-reactions}).}  {Finally, we} provide an overview of reaction rate constants previously calculated for the \ce{CO + H} network  {involving both formaldehyde and methanol} \citep{and11,Goumans2011a,gou11,Song2017} (Section~\ref{sec:prev}). Low-temperature reaction rate constants have been calculated for the first time using instanton theory and serve as an order of magniture estimate implementation in astrochemical models. We will also comment on the possibility to generalize rate constants based only on the type of reaction. 

\section{Computational Details}\label{sec:computation}
Two different levels of theory have been used throughout this study in order to balance the computational cost and chemical accuracy. All calculated activation and reaction energies, as well as the rate constants have been calculated with density functional theory (DFT). In particular, the functional MPWB1K combined with the basis set def2-TZVP has been used. The accuracy of the activation energies or barrier heights is ensured by benchmarking these values to a better level of theory, namely CCSD(T)-F12/VTZ-F12. 

Optimizations of the stationary points, corresponding energies, and spin densities were computed at the MPWB1K/def2-TZVP level \citep{MPWB1K,Weigend2005,Weigend2006}. Geometry optimizations (minima and transition states) were done with DL-FIND \citep{dlfind} in ChemShell \citep{she03,met14}. For the electronic structure computations (energies, gradients, and Hessians) Gaussian 09 \citep{gaussian09} has been employed. SCF cycles were stopped when the convergence, as defined in G09, reached $\num{1E-9}$ Hartree. A pruned (99 590) grid (ultrafine grid) was employed, having 99 radial shells and 590 angular points per shell. 

The MPWB1K functional has been previously benchmarked in order to predict the correct bond dissociation energy of methyl formate (MF), for which accurate results were obtained \citep{Truhlar_mf_2016}. 
Furthermore, MPWB1K was developed to take into account weak interactions such as those found in the pre-reactive complexes treated here. In order to confirm the use of this functional for the current study, single point energy calculations at the RHF-UCCSD(T)-F12/VTZ-F12//MPWB1K/def2-TZVP level \citep{Knowles:1993, Knowles:err, Deegan:1994,Adler:2007, Knizia:2009,Peterson:2008} were carried out and are discussed in Appendix~\ref{benchmark}.

The instanton method based on Feynman path integral theory using the semiclassical approximation was used to compute the reaction rate constants  \citep{lan67,lan69,mil75,col77,cal77,gil77,aff81,col88,han90,ben94,mes95,ric09,kry11,alt11,rom11,rom11b,kry14,ric16}. For a given temperature, it provides the most probable tunnelling path, the instanton, which connects the reactant and product valleys of the potential energy surface. Instanton theory is applicable whenever the temperature is low enough for the instanton to spread out. At higher temperatures, the instanton collapses to a point which renders the theory inapplicable. For most barriershapes this collapse happens at the crossover temperature $T_\text{c}$ \citep{gil87,alv14},
\begin{equation}
 T_\mathrm{c}=\frac{\hbar\ \Omega}{2\pi k_\text{B}} 
\end{equation}
with $\Omega$ being the the absolute value of the imaginary frequency corresponding to the transition mode and $k_\text{B}$ corresponding to Boltzmann's constant. $T_\text{c}$ qualitatively indicates at which temperature the reaction is dominated by tunneling ($T$ $<$ $T_\text{c}$ ) or by the thermal activation ($T$ $>$ $T_\text{c}$).

Instanton paths were optimized via a quasi Newton--Raphson method \citep{rom11,rom11b}. Energies, gradients, and Hessians were provided by Gaussian 09, but instanton optimizations are done in DL-FIND. The instanton path was discretised using 80 images, except for reactions MF3 and MF4 where 158 images were employed at T $\leq$ 100 K and 314 images for MF4 at 75~K.

This study focuses on unimolecular rate constants, \emph{i.e.}, on the Langmuir–Hinshelwood  mechanism. Both reactants are adsorbed on the surface, approach each other via diffusion and form a pre-reactive complex (PRC) on the surface. This PRC can then decay to yield the reaction products via a unimolecular process. It has been shown in the recent literature that often gas-phase calculations of stationary points offer a reasonably accurate approach for representing the very same reactions on an ice surface. This even holds for ices composed of water molecules as typical changes of the activation energy are roughly only 1--2~kJ/mol \citep{Rimola:2014,Song2017,Lamberts:2018}. However, in particular cases, larger energy differences may be found \citep{Lamberts:2017} and to which extent surface molecules may affect the binding orientation is currently unclear. Finally, adsorption on a surface is simulated by keeping the rotational partition function constant between the reactant and transition state. For more information regarding this approach the reader is referred to \citet{Meisner:2017} and \citet{Lamberts:2017B}.

\begin{figure*}
	\centering
	\includegraphics[width=0.71\textwidth]{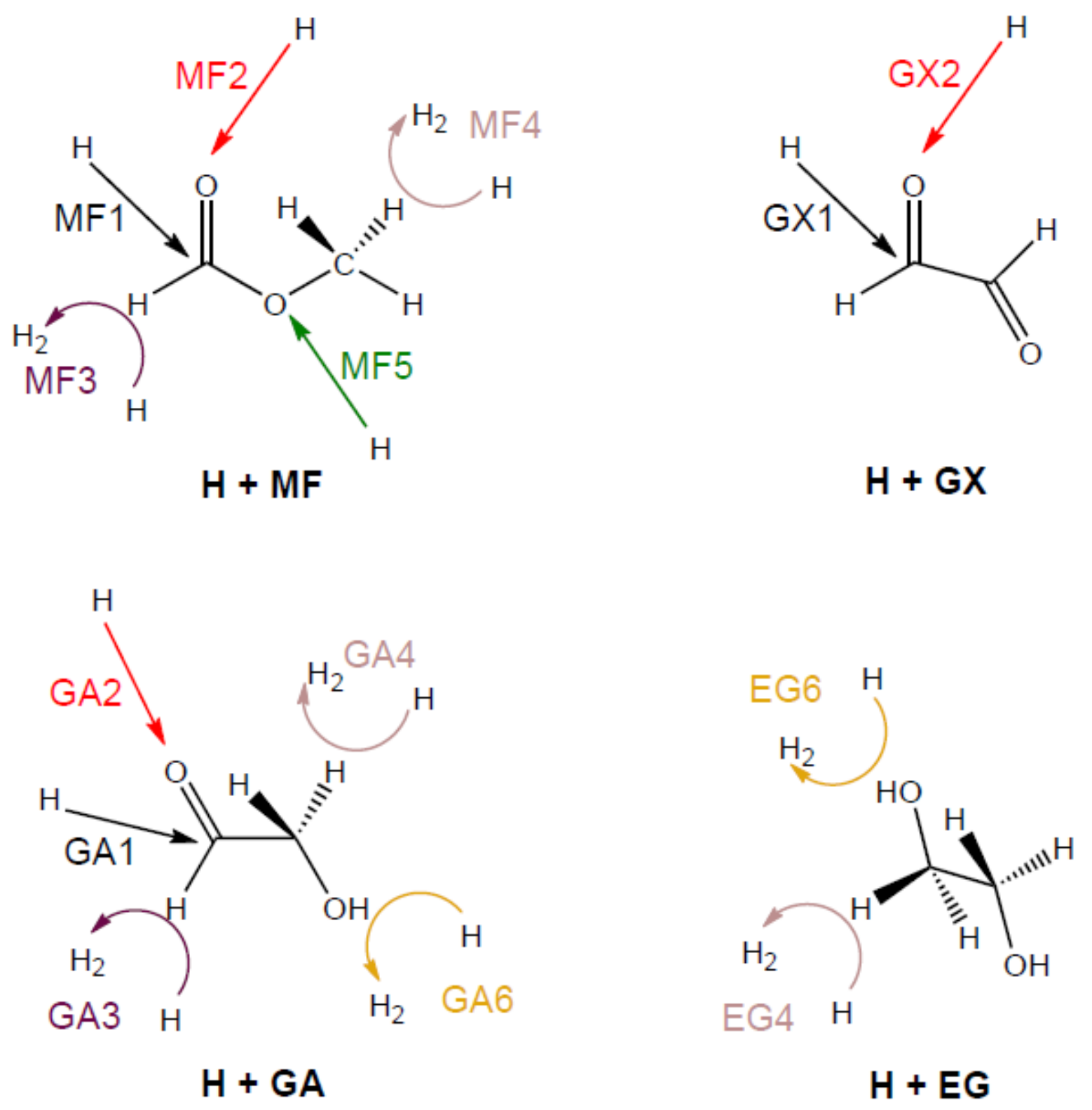}
	\caption{Schematic representation of the reactions shown in section ~\ref{sec:network-new}. MF, GX, GA and EG, 
	 correspond to methylformate, glyoxal, glycoaldehyde, and ethylene glycol, respectively.}
	\label{fig:structures}
\end{figure*}

\section{Results}

 {We simulated a total of 14 reactions revolving around the molecules glyoxal (GX), glycoaldehyde (GA), ethylene glycol (EG), and methyl formate (MF), an additional three reactions where reactions of \ce{CH_nO} fragments with \ce{H2CO} are involved (FAR), and discuss the results in the light of the 6 previously studied reactions with carbon monoxide (CO), formaldehyde (FA), and methanol (ME).}
To structure the analysis, the reactions are labeled according to their type,  {except for the FARn series}:
\begin{enumerate}[1)]
 \item  { H addition to aldehyde carbon}  {-- MF, GX, GA, CO, FA}
 \item  { H addition to aldehyde oxygen}  {-- MF, GX, GA, FA}
 \item  { H abstraction from aldehyde carbon}  {-- MF, GX, GA, FA}
 \item  { H abstraction from methyl group}  {-- MF, EG, ME}
 \item  { H addition to  {etheric} oxygen}  {-- MF}
 \item  { H abstraction from alcohol oxygen}  {-- GA, EG, ME}
\end{enumerate}
{Note that we expect all reactions studied and discussed here to take place in an environment where carbon monoxide, CO, is the main component of the ice mantle. Due to the general weak interactions of this molecule, we expect that the activation energies calculated here in the gas phase will be similar to those in the presence of a CO environment. For instance for the reactions \ce{H + CO} and \ce{H + H2CO} this has been confirmed by \citet{Rimola:2014}. Therefore, the values presented here are thought to be a good representation of the situation in the interstellar medium.}

\subsection{ {Reactions with MF, GX, GA, and EG}}\label{sec:network-new}

Activation energies for the reactions described in this Section and in Section~\ref{sec:FA-reactions} can be found in Table~\ref{tbl:energies}. Crossover temperatures ($T_\text{c}$) as an indication of 
the importance of tunneling, and the calculated low-temperature rate constants are presented as well.
A schematic representation of the reactions with methylformate, glyoxal, glycoaldehyde, and ethylene glycol is given in Figure \ref{fig:structures}.  {The temperature-dependence of the calculated rate constants is depicted in Figs.~\ref{fig:mf}--\ref{fig:IRC}.}
 {The reaction of the hydrogen atom with methylformate has been studied in order to determine if it is an efficient destruction channel. }
 {The addition and abstraction reactions of H with glyoxal, glycoaldehyde, and ethylene glycol serve to study the sequential hydrogenation steps. In this way the same reaction type (see above) can be compared between various molecules, and it can be determined whether or not addition is faster than abstraction. Note that} the reaction abstracting a hydrogen atom from glyoxal (\ref{GX3}) could not be studied, because according to the benchmark study the most accurate value for the activation energy cannot be validated: the difference between the DFT and CCSD(T)-F12 value is too large and moreover multireference effects prevent the CCSD(T)-F12 value from being trusted.

\subsubsection*{Reactions with methylformate (MF)}
\begin{align}
&\ce{H + MF -> CH3OCOH2}  \tag{MF1}\label{MF1}   \\
&\ce{H + MF -> CH3OCHOH} \tag{MF2}\label{MF2} 	 \\
&\ce{H + MF -> CH3OCO + H2} \tag{MF3}\label{MF3} \\ 
&\ce{H + MF -> CH2OCOH + H2} \tag{MF4}\label{MF4}\\
&\ce{H + MF -> HCO + CH3OH} \tag{MF5}\label{MF5} 
\end{align} 

\subsubsection*{Reactions with glyoxal (GX)}
\begin{align}
&\ce{H + GX ->  (H2CO)CHO } \tag{GX1}\label{GX1}  \\
&\ce{H + GX ->  (HCOH)CHO } \tag{GX2}\label{GX2}  \\
&\ce{H + GX ->  (CO)CHO + H2} \tag{GX3}\label{GX3} 
\end{align}

\subsubsection*{Reactions with glycoaldehyde (GA)}
\begin{align}
&\ce{H + GA ->  (H2CO)CH2OH } \tag{GA1}\label{GA1}  \\
&\ce{H + GA ->  (HCOH)CH2OH } \tag{GA2}\label{GA2}  \\
&\ce{H + GA ->  (CO)CH2OH + H2 } \tag{GA3}\label{GA3}  \\
&\ce{H + GA ->  (HCO)CHOH + H2 } \tag{GA4}\label{GA4}  \\
&\ce{H + GA ->  (HCO)CH2O + H2 } \tag{GA6}\label{GA6}  
\end{align} 

\subsubsection*{Reactions with ethylene glycol (EG)}
\begin{align}
&\ce{H + EG -> (HOCH)CH2OH + H2 } \tag{EG4}\label{EG4}   \\
&\ce{H + EG -> (OCH2)CH2OH + H2 } \tag{EG6}\label{EG6}  
\end{align}

\begin{table}
\centering
  \caption{Activation energies with respect to the pre-reactive complexes computed at the MPWB1K/def2-TZVP level with ($\Delta E^{0,\ddagger}$) and without ($\Delta E^{\ddagger}$) ZPE correction. Crossover temperatures ($T_\text{c}$) and rate constansts ($k$ at 75~K unless indicated otherwise) are also included.
  \label{tbl:energies}
    }
  \begin{tabular}{lrrrl}
      \hline											
	&	$\Delta E^{\ddagger}$	&	$\Delta E^{0,\ddagger}$	&	$T_\text{c}$	&	$k$  (75 K)	\\
	& 	{(kJ~mol$^{-1}$)} 	& {(kJ~mol$^{-1}$)}	 	& (K)			& s$^{-1}$ \\
    \hline											
 {H + MF}	&		&			&		&		\\
\ref{MF1}	&	38.1	&	41.2	&		262.2	&	\num{3.6E-01}	\\
\ref{MF2}	&	59.0	&	59.9	&		400.4	&	\num{1.7E+00}	\\
\ref{MF3}	&	46.6	&	38.1	&		365.8	&	\num{3.4E+00}$^{a}$	\\
\ref{MF4}	&	51.2	&	42.8	&		345.1	&	\num{1.1E-01}	\\
\ref{MF5}	&	146.0	&	149.9	&		496.7	&	\num{3.8E-33}	\\
& & & & \\
 {H + GX}	&		&	&			&		\\
\ref{GX1}	&	15.1	&	15.1	&		179.9	&	\num{9.6E+06}	\\
\ref{GX2}	&	29.8	&	31.7	&		298.9	&	\num{1.8E+03}	\\
	&		&		&			&		\\
& & & & \\
 {H + GA}	&		&		&			&		\\
\ref{GA1}	&	19.0	&	20.8	&		203.4	&	\num{2.8E+05}	\\
\ref{GA2}	&	38.5	&	39.8	&		342.5	&	\num{2.8E+02}	\\
\ref{GA3}	&	24.3	&	14.6	&		317.6	&	\num{6.8E+07}	\\
\ref{GA4}	&	27.3	&	20.6	&		333.5	&	\num{2.6E+04}	\\
\ref{GA6}	&	55.8	&	46.2	&		405.2	&	\num{9.6E-01}	\\
	&		&		&			&		\\
 {H + EG}	&		&	&			&		\\
\ref{EG4}	&	28.4	&	19.3	&		303.3	&	\num{3.5E+06}	\\
\ref{EG6}	&	54.1	&	42.2	&		406.3	&	\num{2.1E3}$^{b}$	\\
	&		&		&		&		\\
 {FARn}	&		&	&						&		\\
\ref{FAR1}	&	30.6	&	22.2	&		336.6	&	\num{2.0E3}$^{c}$	\\
\ref{FAR2}	&	44.8	&	48.5	&		151.9	&	\num{4.6E-9}$^{d}$	\\
\ref{FAR3}	&	19.9	&	24.5	&		58.3	&	\num{3.9E-11}$^{c}$	\\
   \hline											
\multicolumn{5}{l}{$^{a}$ at 80~K}\\
\multicolumn{5}{l}{$^{b}$ at 90~K}\\
\multicolumn{5}{l}{$^{c}$ at 50~K}\\
\multicolumn{5}{l}{$^{d}$ at 65~K}
\end{tabular}
\end{table}

\subsection{Reactions between FA and \ce{CH_nO}} \label{sec:FA-reactions}
 {Although the COMs discussed above have been proposed to be formed mainly through radical-radical reactions, reactions between a neutral and radical species may also lead to the formation of a C--C or C--O bond. The reactions between \ce{H2CO} and \ce{CH3O} or \ce{HCO} \citep{but17} are therefore studied as well in order to compare their efficiency to other radical-neutral reactions as well as to fast barrierless radical-radical reactions.}

\begin{figure}
        \centering
        \includegraphics[width=0.44\textwidth]{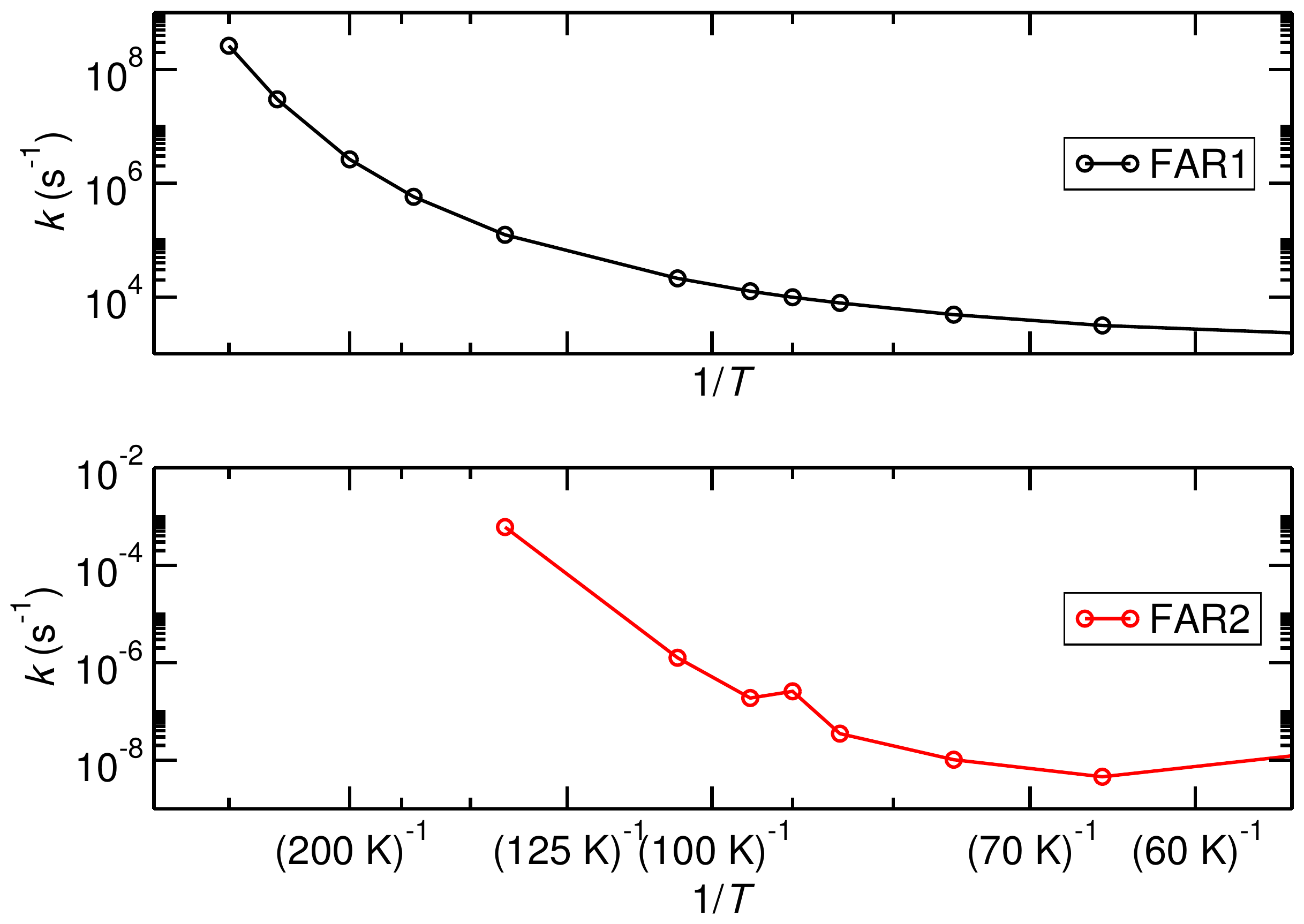}
        \caption{Unimolecular rate constants (in $s^{-1}$) calculated with instanton theory for the
                  {reactions of formaldehyde with \ce{CH_nO} fragments (FARn)}.}
        \label{fig:fa}
\end{figure}

Reactions \ref{FAR1} and~\ref{FAR2}  {are in direct competition with each other}, see also Fig.~\ref{fig:fa}.
\begin{align}
&\ce{H2CO + CH3O -> CH3OH + HCO} \tag{FAR1}\label{FAR1}   \\
&\ce{H2CO + CH3O -> CH3OCH2O} \tag{FAR2}\label{FAR2} \\
&\ce{H2CO + HCO -> (HCO)CH2O} \tag{FAR3}\label{FAR3} 
\end{align}

\subsection{Reactions with CO, FA, and ME}\label{sec:prev}

 {Prior to discussing hydrogen transfer reactions in COMs}, this Section summarizes previous theoretical studies related to the \ce{H + CO} reaction network for cases where calcualtions have also been performed with instanton theory. The main results, in terms of activation energy and reaction rate constant from those studies are listed in Table~\ref{tbl:lit}.

\begin{table}
 \centering
 \caption{Activation energies including ZPE ($\Delta E^{0,\ddagger}$) and unimolecular rate constants ($k$) obtained from literature values.}\label{tbl:lit}
 \begin{tabular}{lrrl}
 \hline
      & 	$\Delta E^{0,\ddagger}$	& 	$k$ \\
    & 		(kJ/mol)	& s$^{-1}$ \\
      \hline
 {H + CO} & & & \\
 \ref{CO1}	& 12.4 + $\sim$1.2$^{a}$& \num{2.1E5} at 5~K			& [1]	\\
 & & & \\
 {H + FA} & & & \\ 
 \ref{FA1}	& 15.8 - 17.9 		& \num{1.5E5} - \num{2.0E6} at 70~K 	& [2]	\\
 \ref{FA2}	& 43.3 - 47.1		& \num{4.0E1} - \num{9.0E1} at 75~K 	& [2]	\\
 \ref{FA3}	& 20.5 - 25.2		& \num{4.0E5} - \num{1.0E6} at 70~K 	& [2]	\\
  & & & \\
 {H + ME} & & & \\
 \ref{ME4}	& 30.2			& --					& [3]	\\
 \ref{ME6}	& 46.4			& -- 					& [3]	\\ 
\hline
\multicolumn{3}{l}{$^{a}$ ZPE calculated in this work (CCSD(T)-F12/VTZ-F12)}\\
\multicolumn{3}{l}{[1] \cite{Andersson2011} [2]  \cite{Song2017}}\\
\multicolumn{3}{l}{[3] \cite{Goumans2011a}}
\end{tabular} 
\end{table}

\begin{figure}
	\centering
	\includegraphics[width=0.44\textwidth]{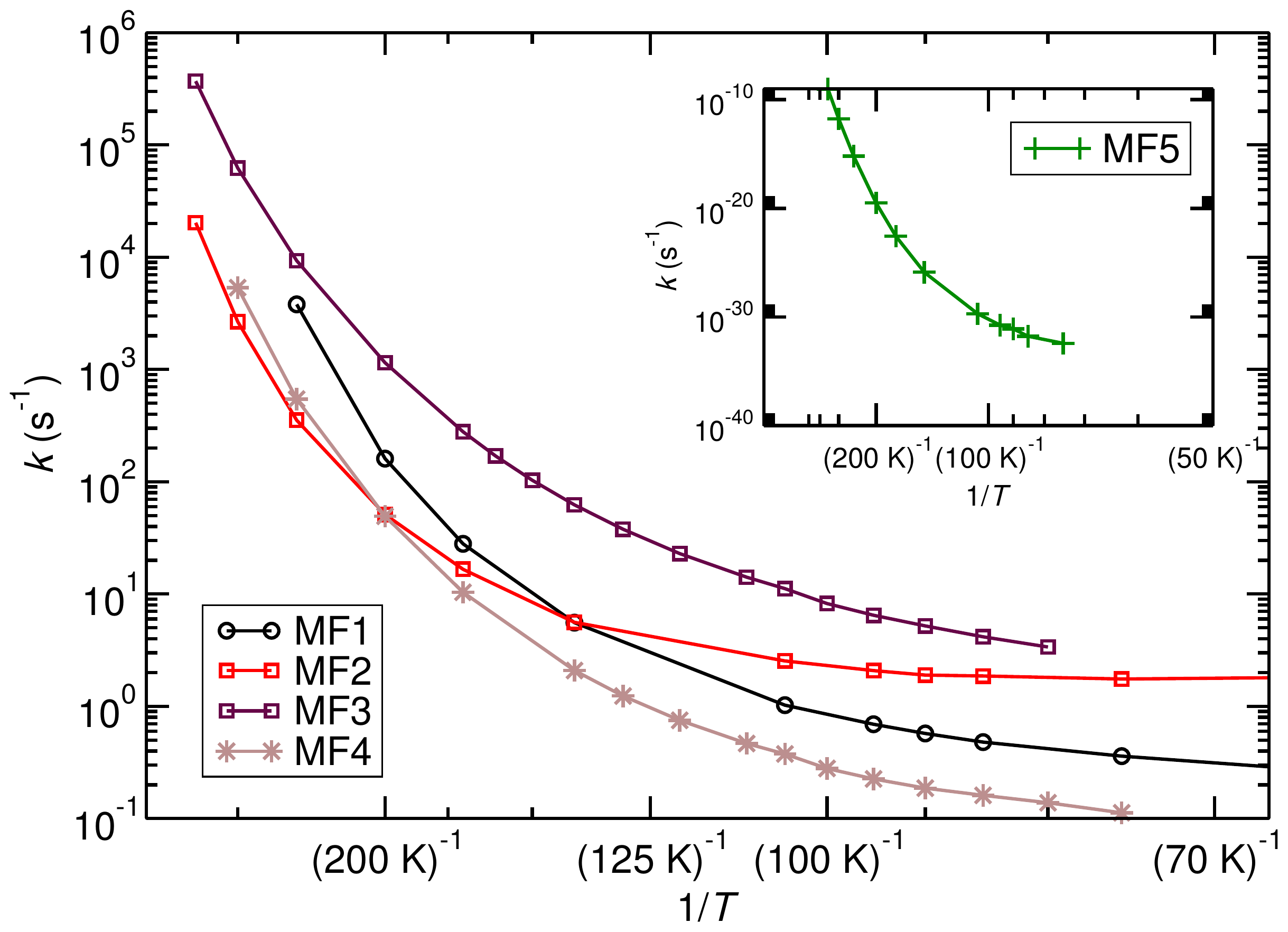}
	\caption{Unimolecular rate constants (in $s^{-1}$) calculated with instanton theory for the  {methylformate (MF)} + H reactions.}
	\label{fig:mf}
\end{figure}

\begin{figure}
        \centering
        \includegraphics[width=0.44\textwidth]{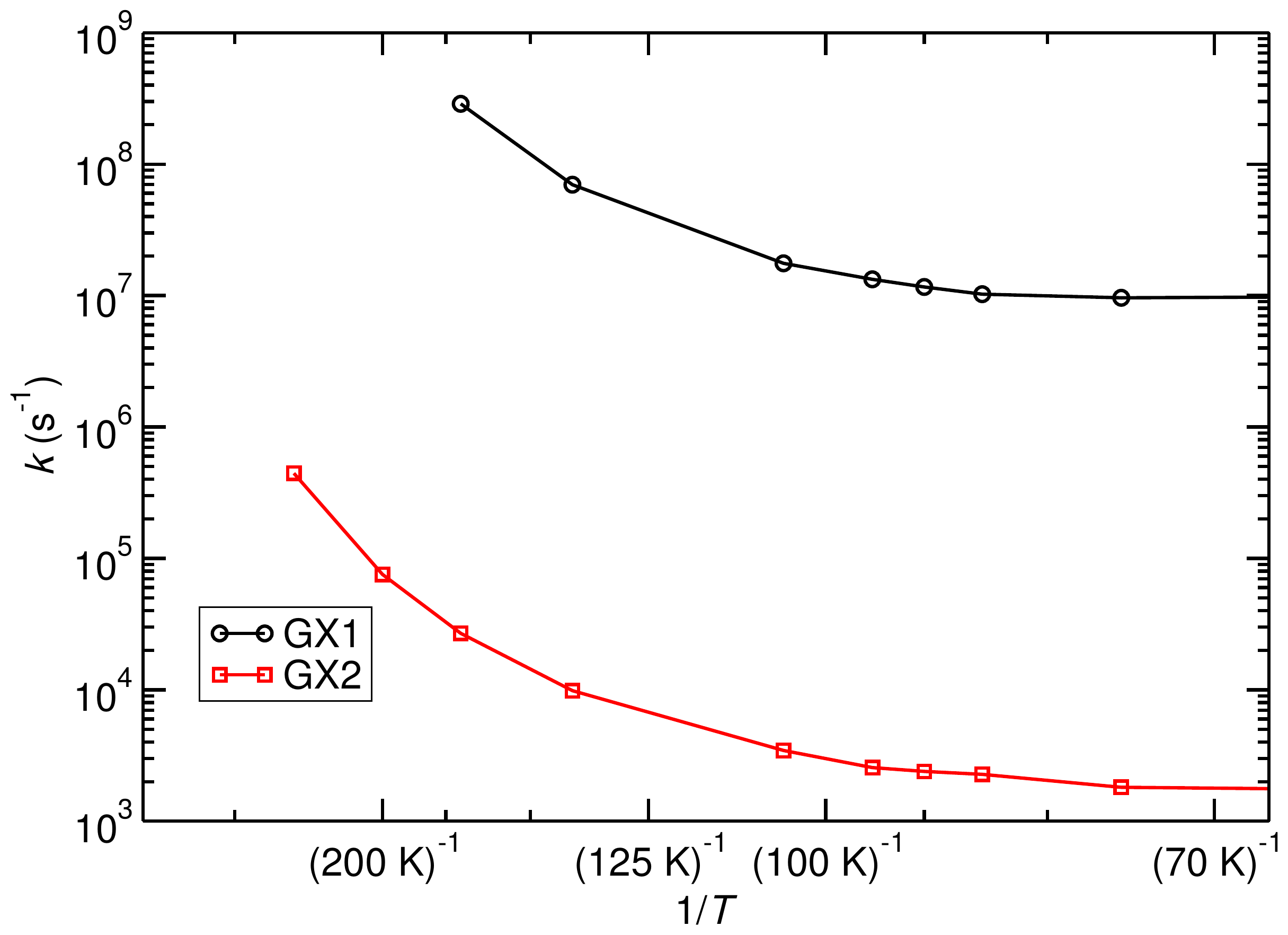}
        \caption{Unimolecular rate constants (in $s^{-1}$) calculated with instanton theory for the  {glyoxal (GX)} + H reactions.}
        \label{fig:gx}
\end{figure}

\begin{figure}
        \centering
        \includegraphics[width=0.44\textwidth]{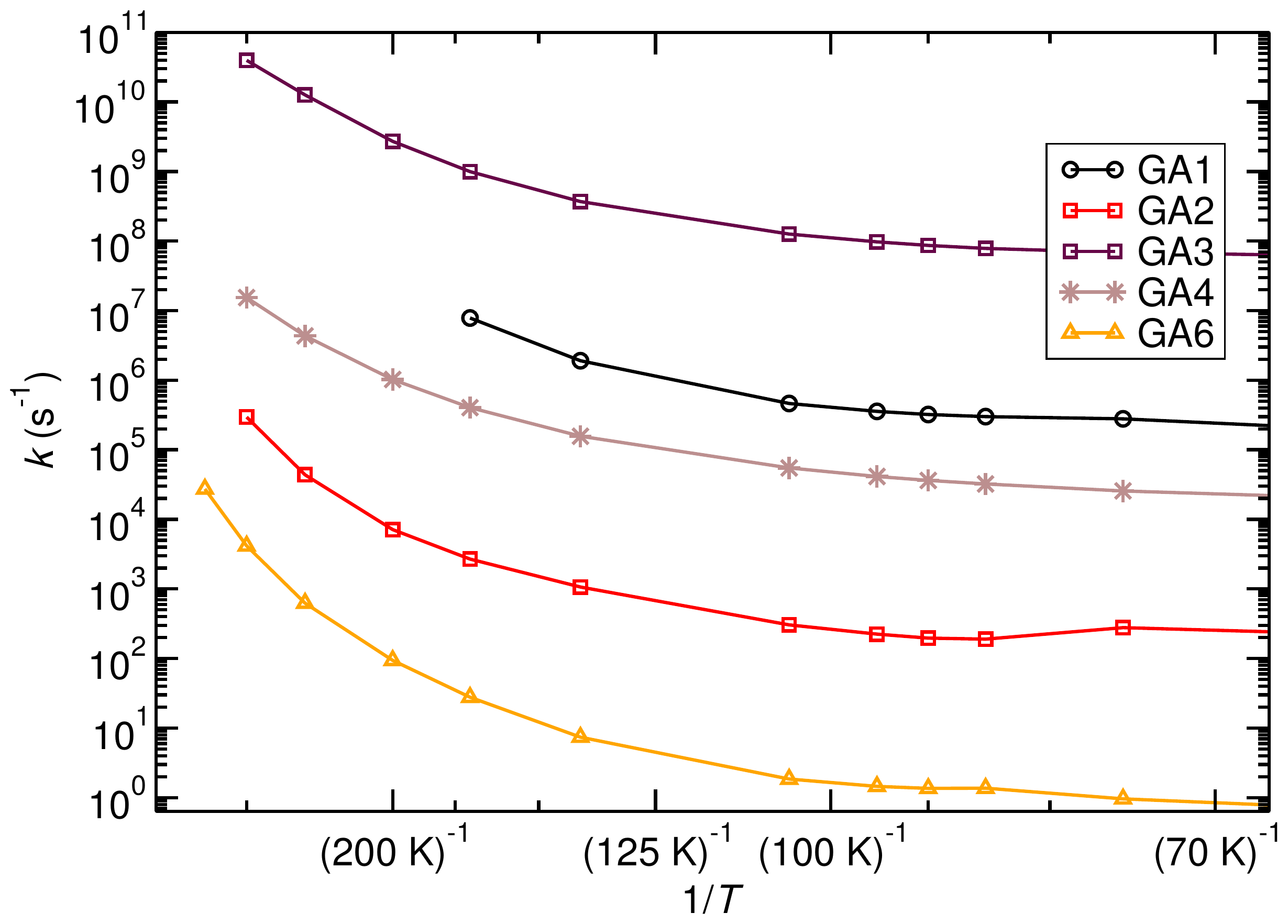}
        \caption{Unimolecular rate constants (in $s^{-1}$) calculated with instanton theory for the  {glycoaldehyde (GA)} + H reactions.}
        \label{fig:ga}
\end{figure}

\begin{figure}
        \centering
        \includegraphics[width=0.44\textwidth]{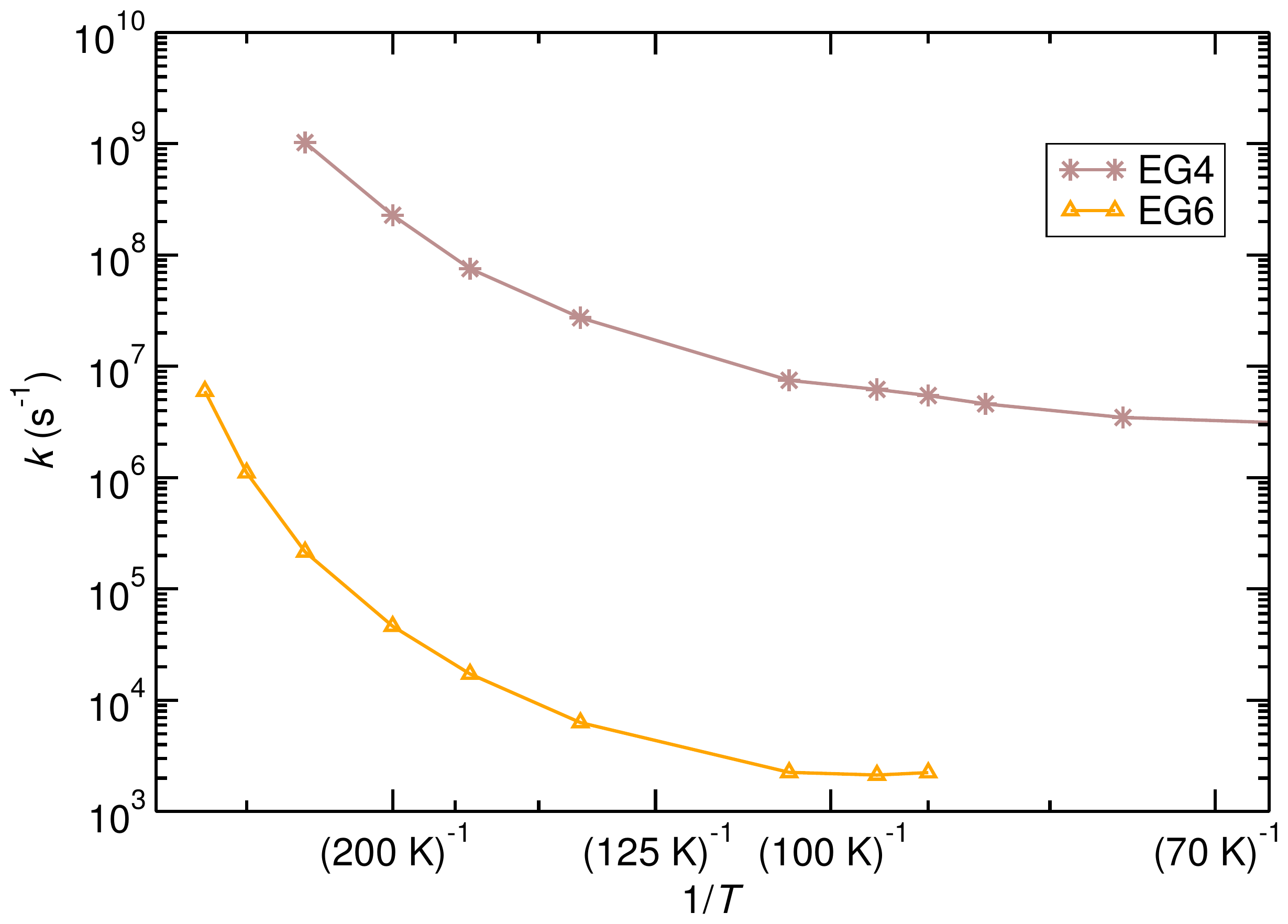}
        \caption{Unimolecular rate constants (in $s^{-1}$) calculated with instanton theory for the  {ethylene glycol (EG)} + H reactions.}
        \label{fig:eg}
\end{figure}

\begin{figure}
        \centering
        \includegraphics[width=0.44\textwidth]{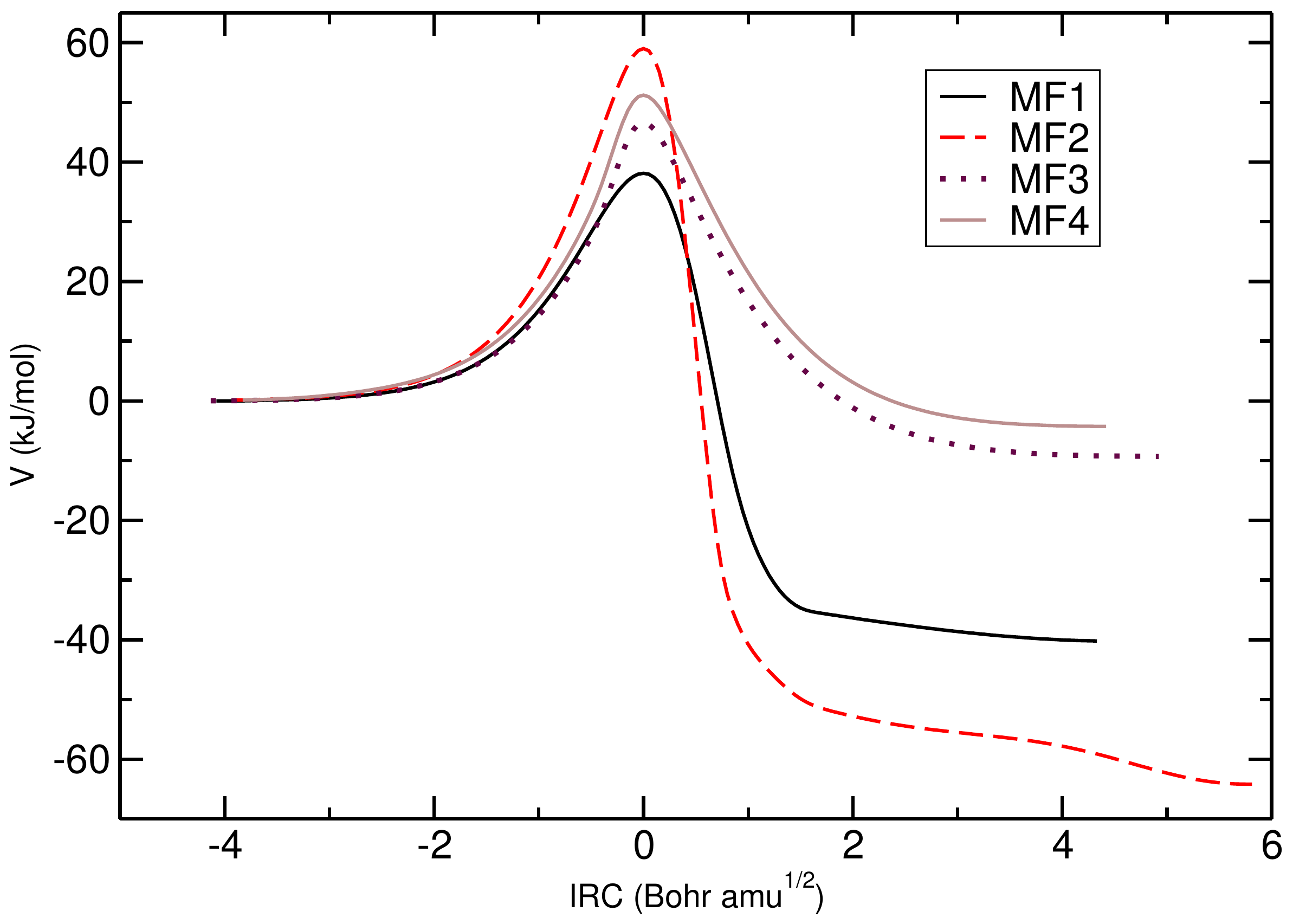}
        \caption{Intrinsic reaction coordinates for reactions \ref{MF1} - \ref{MF4}.}
        \label{fig:IRC}
\end{figure}

\clearpage

\subsubsection*{Reaction with CO}
Unimolecular rate constants for the \ce{H + CO} system have been theoretically calculated by \citet{Andersson2011}, using a PES previously obtainded by \citet{Keller1996}:
\begin{align}
&\ce{H + CO -> HCO}  \tag{CO1}\label{CO1}
\end{align} 

\subsubsection*{Reactions with FA}

The reaction of H and \ce{H2CO} (\ce{H + FA}) has been theoretically studied by both \citet{gou11b} and \citet{Song2017}. 
\begin{align}
&\ce{H + FA -> CH3O}  \tag{FA1}\label{FA1}    \\
&\ce{H + FA -> CH2OH} \tag{FA2}\label{FA2}    \\
&\ce{H + FA -> H2 + HCO} \tag{FA3}\label{FA3} 
\end{align}

\subsubsection*{Reactions with ME}
The abstraction of H on the methanol (ME) mehtyl group has been studied theoretically by \citet{Goumans2011a}. 
\begin{align}
&\ce{H + ME -> CH2OH + H2}  \tag{ME4}\label{ME4} \\ 
&\ce{H + ME -> CH3O + H2}  \tag{ME6}\label{ME6}   
\end{align}

\section{Discussion}

\subsection{ {Hydrogen addition and abstraction reactions}}

 {For most of the reactions, the rate constants level off at a given temperature, especially below 80~K.}

The reactions of the hydrogen atom with methylformate overall have the highest activation energies, indicating that MF is quite stable with respect to attack by a H radical. Reaction \ref{MF5} is a special case, where a C-O bond is being broken while an O-H bond is formed, which is most likely the reason for the corresponding activation energy, or high barrier.

Comparing between the various reaction types, Figures~\ref{fig:mf} and ~\ref{fig:ga} show that the $\ce{H}$ abstractions from the H of the HC=O group (type 3) occur with high reaction rate constants. The $\ce{H}$ abstraction from the -OH group (type 6), on the other hand, appears to be very unfavourable with barriers larger than $\sim$50~kJ/mol (Figures~\ref{fig:ga} and ~\ref{fig:eg}). This is consistent with previous results obtained for the reaction between hydrogen and methanol. The barrier  {between reactions \ref{ME4} and \ref{ME6}} differs by 16 kJ/mol, in line with the experimental work of \citet{chu16} and \citet{Nagaoka:2007}. 

Generalizing reaction types 1, 2, and 4 is not trivial. The activation energies for type 1 are always lower than those for type 2, when compared within the same molecule (FA, MF, GX, and GA). For instance the formation of \ce{CH3O} is preferred over the formation of \ce{CH2OH}, contrary to the findings of \citet{but15}, but in line with those of \citet{chu16}. The rate constants for type 1 are indeed higher than those for type 2 for reactions with FA, GX, and GA, but MF is a special case. Similarly for type 4, where the activation energies are higher than for type 2 and consequently the rate constants are lower for GA, but again reaction \ref{MF2} deviates. At temperatures below 200~K the reaction rate constant for \ref{MF2} crosses first that of \ref{MF4} and later that of \ref{MF1} even though the barrier is higher. The origin of this behavior lies in the barrier width. Tunneling namely depends both on the barrier height and width as well as on the effective mass of the system. The narrower the barrier at low-energy incidence, the more tunneling may be expected. This can be visualized with the help of intrinsic reaction coordinates (IRCs). The IRC curves are presented in Fig.~\ref{fig:IRC}, note that these do not include ZPE corrections and therefore relate to the barrier height $\Delta E^{\ddagger}$ of Table~\ref{tbl:energies}.

 {Finally, it cannot be said that in general addition is more efficient than abstraction or vice versa, \emph{e.g.}, compare reaction types 1 and 2 against 3, 4, and 6.}

\subsection{ {Reactions between FA and \ce{CH_nO}}}

 {At decreasing temperatures tunneling dominates a reaction more and more. This can explain the large difference between the low temperature value for the rate constant of reaction \ref{FAR2} compared to \ref{FAR1}. For \ref{FAR2} a C--O bond is formed and as heavy-atom tunneling is less efficient than hydrogen atom tunneling, the rate constant is much lower than what would be expected from the barrier height only (compare for instance \ref{MF4}, \ref{GA6}, and \ref{FAR2}).  
Similarly for the reaction of FA with \ce{HCO}, where a C--C bond is formed (\ref{FAR3}), again the low-temperature rate constant is very low. Note also the lower values for the crossover temperatures of \ref{FAR2} and \ref{FAR3} compared to the hydrogen transfer reactions, indicating that tunneling also sets in at lower temperatures. Comparing the rate constant to the typical value assumed for radical-radical barrierless reactions, $\sim10^{12}$ s$^{-1}$, it is clear that these reactions are much less likely to contribute to COM formation. On that note we do wish to stress, however, to keep in mind that although radical-radical reactions may be able to proceed without a barrier, this does not mean that all reaction pathways are open, see for instance \citet{Lamberts:2018}}.

\section{Astrochemically relevant conclusions}
Unimolecular reaction rate constants have been calculated and are provided for hydrogen addition and abstraction reactions from methylformate, glyoxal, glycoaldehyde, and ethylene glycol and are thus available to be implemented in both rate-equation and kinetic Monte Carlo models aimed at studying the formation of COMs at low temperatures.

Our results are generally in agreement with experimental work, although some discrepancies exist on the efficiency of specific reaction paths, such as the formation of \ce{CH2OH} or \ce{CH3O} after hydrogen abstraction from methanol, which impacts on the ease of methylformate formation (for which \ce{CH3O} is needed) or ethylene glycol formation (for which \ce{CH2OH} is required). A microscopic model aiming to reproduce experiments may be able to provide a clear picture of how the reactions are intertwined with each other. 

The reaction \ce{H + GX -> (CO)CHO + H2} could not be studied and thus deserves further attention.

We found that one cannot predict average rate constants solely based on the type of the reaction. The spread in the low-temperature rate constant can be roughly 7 orders of magnitude for a single reaction type (\emph{e.g.}, hydrogen addition to an aldehyde carbon) showing a strong dependence on the other functional groups that are attached to the carbon backbone. 

Within a single molecule, on the other hand, one can loosely say that hydrogen abstraction from an aldehyde group is faster than hydrogen addition to the same carbon. Both of these have a rate constant that is larger than hydrogen abstraction from a methyl group.

Care should be taken with extrapolating rate constants based on the height of the barrier alone, as calculations show that reactions with narrow barriers can have rate constants at low temperature that are higher than those with a lower activation energy.

Reactions that include the breakage or formation of a bond between two heavy atoms generally have low-temperature rate constants that are much lower than those for hydrogen addition or abstraction reactions as a result of the low efficiency of tunneling when heavy atoms are involved.

\section*{Acknowledgements}
The authors acknowledges support for computer time by the state of Baden-W\"{u}rttemberg through bwHPC and the Germany Research Foundation (DFG) through grant no. INST 40/467-1FUGG and SFB 716/C.6.
This project was financially supported by the European Union's Horizon 2020 research and innovation programme (grant agreement No. 646717, TUNNELCHEM), the Alexander von Humboldt Foundation, the Netherlands Organisation for Scientific Research (NWO) via a VENI fellowship (722.017.008) and the COST Action CM1401 via an STSM travel grant.

\bibliographystyle{mnras}
\bibliography{text,rsc} 

\appendix

\section{Benchmark calculations}\label{benchmark}

RHF-UCCSD(T)-F12/VTZ-F12//MPWB1K/def2-TZVP single-point energy calculations were performed in order to check if MPWB1K provides a suitable description of the energy landscape for the reactions studied here.
In general, the CCSD(T)-F12 method can be seen as the gold standard for obtaining relative energies for systems that are well-described by a single reference wavefunction. This is typically assumed to be the case when the so-called T1 and D1 diagnostics are smaller than the commonly used threshold values ($T1 \leq 0.02$ and $D1 \leq 0.05$) \citep{Lee2003}. Here, this is the case for reactions MF2, MF4, GX1, GA1, GA4, and EG4. These reactions are included in Table~\ref{tbl:benchmark} and the deviation in the activation energy ranges between 0.4 and 3.3 kJ/mol, \emph{i.e.}, within chemical accuracy.

Furthermore, the extent of the multireference character for reaction type 2 (MF2, GX2 and GA2) was tested via MRCI-F12/VTZ-F12//MPWB1K/def2-TZVP calculations \citep{Shiozaki:2011,Shiozaki:2011a,Peterson2008} for a reaction of the same type, but with a smaller reactant: \ce{H + H2CO -> CH2OH}. These single-point energy calculations indicate that the reaction does not have a large multireference character. Firstly, the CI coefficients for the reference wavefuntion of the transition state structure correspond to 0.934, -0.124, 0.074 and -0.051. In addition, the activation energies at DFT, CCSD(T)-F12 and MRCI-F12 level are similar (38.8, 40.6 and 36.4 kJ/mol, respectively). Therefore, here the CCSD(T)-F12 method is considered to be a reasonable reference method for these specific three reactions as well, \emph{i.e.}, for a H addition to an aldehyde oxygen.

The MPWB1K functional has been shown to provide a good description for 9 out of the 17 reactions dealt with here, with reaction types 1 to 4 being included in this benchmark. Therefore, we assume that the other 8 reactions, including reaction type 5 and 6, can also be described with the same functional and basis set combination. As a double-check we have tested several functionals suggested by \citet{Truhlar_mf_2016} to make sure that the activation energies obtained are of the correct magnitude, see Table~\ref{tbl:dft_comp}.

\begin{table}
 \centering
 \caption{Activation energies without zero-point energy correction  ($\Delta E^{\ddagger}$) with respect to the separated reactants (in kJ~mol$^{-1}$) computed at the MPWB1K/def2-TZVP level (DFT) and RHF-UCCSD(T)-F12/VTZ-F12//MPWB1K/def2-TZVP (CC)}\label{tbl:benchmark}
 \begin{tabular}{lll}
\hline
 & $\Delta E^{\ddagger}$ DFT	& $\Delta E^{\ddagger}$ CC \\
	\hline
MF2	&	58.8	&	59.2			\\
MF3	&	46.5	&	\textit{48.3}$^{a}$	\\
MF4	&	51.1	&	52.8			\\
GX1	&	14.7	&	15.4			\\
GX2	&	29.7	&	\textit{31.9}$^{a}$	\\
GA1	&	18.5	&	17.5			\\
GA2	&	38.2	&	\textit{37.7}$^{a}$	\\
GA4	&	26.5	&	29.5			\\
EG4	&	27.9	&	31.2			\\
\hline
\multicolumn{3}{l}{$^{a}$ Single-reference character}\\ 
\multicolumn{3}{l}{confirmed, see text}
\end{tabular}
\end{table}

\begin{table*}
 \centering
 \caption{Activation energies without zero-point energy correction ($\Delta E^{\ddagger}$) with respect to the separated reactants (in kJ~mol$^{-1}$) computed with several functionals and the def2-TZVP basis set}\label{tbl:dft_comp}
 \begin{tabular}{lllllll}
\hline													
	&	MPWB1K	&	M06-2X	&	MPW1B95	&	MN12-SX	&	N12-SX	&	SOGGA11-X	\\
\hline													
 {H + MF}	&		&		&		&		&		&		\\
MF1	&	38.0	&	39.2	&	35.0	&	33.7	&	43.1	&	39.1	\\
MF2	&	58.8	&	57.1	&	52.0	&	56.7	&	59.3	&	61.3	\\
MF3	&	46.5	&	50.5	&	38.0	&	43.3	&	43.5	&	46.2	\\
MF4	&	51.1	&	55.1	&	44.3	&	50.1	&	48.7	&	52.5	\\
MF5	&	145.8	&	138.0	&	131.5	&	142.3	&	137.6	&	148.1	\\
 {H + GX}	&		&		&		&		&		&		\\
GX1	&	14.7	&	17.0	&	12.9	&	8.3	&	21.9	&	15.7	\\
GX2	&	29.7	&	33.1	&	24.5	&	24.3	&	32.2	&	30.1	\\
 {H + GA}	&		&		&		&		&		&		\\
GA1	&	18.5	&	18.0	&	16.8	&	11.3	&	27.4	&	20.0	\\
GA2	&	38.2	&	36.9	&	32.4	&	35.8	&	42.3	&	39.8	\\
GA3	&	23.8	&	28.4	&	16.1	&	20.9	&	24.8	&	23.1	\\
GA4	&	26.5	&	32.4	&	20.1	&	26.7	&	28.3	&	27.0	\\
GA6	&	55.4	&	59.3	&	45.7	&	50.5	&	51.9	&	52.8	\\
 {H + EG}	&		&		&		&		&		&		\\
EG4	&	27.9	&	33.2	&	21.0	&	29.0	&	28.1	&	29.2	\\
EG6	&	53.5	&	57.8	&	44.4	&	49.6	&	50.7	&	51.3	\\
   \hline													
\end{tabular}
\end{table*}

\section{Rate Constants}

Tables~\ref{tbl:k_MF}-\ref{tbl:k_FA} give the values for the unimolecular reaction rate constants as calculated with instanton theory and corresponding to Figs.~\ref{fig:mf}-\ref{fig:eg} and~\ref{fig:fa} of the main manuscript.

\begin{table*}
 \centering
 \caption{Unimolecular reaction rate constants ($k$ (s$^{-1}$)) for reaction MF + H. The instanton path was discretised using 80 images.}\label{tbl:k_MF}
 \begin{tabular}{llllll}
 \hline											
\hline											
T(K)	&	MF1		&	MF2		&	MF3		&		MF4	&	MF5		\\
\hline											
75	&	3.60E-01	&	1.75E+00	&			&	1.13E-01$^{b}$	&	3.77E-33	\\
80	&			&			&	3.37E+00$^{a}$	&	1.38E-01$^{a}$	&			\\
85	&	4.80E-01	&	1.86E+00	&	4.16E+00$^{a}$	&	1.61E-01$^{a}$	&	1.70E-32	\\
90	&	5.72E-01	&	1.90E+00	&	5.20E+00$^{a}$	&	1.87E-01$^{a}$	&	8.08E-32	\\
95	&	6.91E-01	&	2.08E+00	&	6.47E+00$^{a}$	&	2.25E-01$^{a}$	&	1.78E-31	\\
100	&			&			&	8.23E+00$^{a}$	&	2.79E-01	&			\\
105	&	1.02E+00	&	2.54E+00	&	1.12E+01	&	3.76E-01	&	1.95E-30	\\
110	&			&			&	1.41E+01	&	4.66E-01	&			\\
120	&			&			&	2.29E+01	&	7.50E-01	&			\\
130	&			&			&	3.77E+01	&	1.24E+00	&			\\
140	&	5.56E+00	&	5.60E+00	&	6.22E+01	&	2.08E+00	&	1.31E-26	\\
150	&			&			&	1.03E+02	&			&			\\
160	&			&			&	1.70E+02	&			&			\\
170	&	2.80E+01	&	1.66E+01	&	2.78E+02	&	1.04E+01	&	2.73E-23	\\
200	&	1.61E+02	&	5.11E+01	&	1.15E+03	&	4.93E+01	&	3.13E-20	\\
250	&	3.80E+03	&	3.53E+02	&	9.31E+03	&	5.45E+02	&	6.43E-16	\\
300	&			&	2.66E+03	&	6.22E+04	&	5.35E+03	&	1.72E-12	\\
350	&			&	2.03E+04	&	3.70E+05	&			&	9.58E-10	\\
\hline											
\multicolumn{3}{l}{$^{a}$ 158 images}\\ 
\multicolumn{3}{l}{$^{b}$ 314 images}
\end{tabular}
\end{table*}

\begin{table*}
 \centering
 \caption{Unimolecular reaction rate constants ($k$ (s$^{-1}$)) for reaction GX + H. The instanton path was discretised using 80 images.}\label{tbl:k_GX}
 \begin{tabular}{lll}
 \hline					
T(K)	&	GX1		&	GX2	\\
\hline					
75	&	9.62E+06	&	1.81E+03	\\
85	&	1.02E+07	&	2.27E+03	\\
90	&	1.16E+07	&	2.39E+03	\\
95	&	1.33E+07	&	2.56E+03	\\
105	&	1.76E+07	&	3.45E+03	\\
140	&	6.97E+07	&	9.85E+03	\\
170	&	2.88E+08	&	2.69E+04	\\
200	&			&	7.55E+04	\\
250	&			&	4.44E+05	\\
\hline					
\end{tabular}
\end{table*}

\begin{table*}
 \centering
 \caption{Unimolecular reaction rate constants ($k$ (s$^{-1}$)) for reaction GA + H. The instanton path was discretised using 80 images.}\label{tbl:k_GA}
 \begin{tabular}{llllll}
\hline											
T(K)	&	GA1		&	GA2		&	GA3		&	GA4		&	GA6	\\
\hline											
75	&	2.78E+05	&	2.77E+02	&	6.83E+07	&	2.56E+04	&	9.61E-01	\\
85	&	3.00E+05	&	1.90E+02	&	7.84E+07	&	3.23E+04	&	1.37E+00	\\
90	&	3.22E+05	&	1.97E+02	&	8.66E+07	&	3.63E+04	&	1.36E+00	\\
95	&	3.56E+05	&	2.24E+02	&	9.74E+07	&	4.13E+04	&	1.45E+00	\\
105	&	4.62E+05	&	3.04E+02	&	1.26E+08	&	5.48E+04	&	1.85E+00	\\
140	&	1.91E+06	&	1.07E+03	&	3.70E+08	&	1.57E+05	&	7.40E+00	\\
170	&	7.83E+06	&	2.68E+03	&	9.96E+08	&	4.04E+05	&	2.77E+01	\\
200	&			&	7.12E+03	&	2.72E+09	&	1.02E+06	&	9.41E+01	\\
250	&			&	4.37E+04	&	1.26E+10	&	4.34E+06	&	6.25E+02	\\
300	&			&	2.96E+05	&	3.96E+10	&	1.54E+07	&	4.16E+03	\\
350	&			&			&			&			&	2.74E+04	\\
\hline											
\end{tabular}
\end{table*}

\begin{table*}
 \centering
 \caption{Unimolecular reaction rate constants ($k$ (s$^{-1}$)) for reaction EG + H. The instanton path was discretised using 80 images.}\label{tbl:k_EG}
 \begin{tabular}{lll}
\hline					
T(K)	&	EG4		&	EG6	\\
\hline					
75	&	3.47E+06	&		\\
85	&	4.60E+06	&		\\
90	&	5.46E+06	&	2.24E+03	\\
95	&	6.20E+06	&	2.13E+03	\\
105	&	7.48E+06	&	2.25E+03	\\
140	&	2.73E+07	&	6.30E+03	\\
170	&	7.54E+07	&	1.72E+04	\\
200	&	2.28E+08	&	4.63E+04	\\
250	&	1.02E+09	&	2.15E+05	\\
300	&			&	1.11E+06	\\
350	&			&	5.98E+06	\\
\hline					
\end{tabular}
\end{table*}

\begin{table*}
 \centering
 \caption{Unimolecular reaction rate constants ($k$ (s$^{-1}$)) for FA reactions. The instanton path was discretised using 80 images.}\label{tbl:k_FA}
 \begin{tabular}{lll}
\hline					
\hline					
T(K)	&	FAR1		&	FAR2	\\
\hline					
50	&	1.99E+03	&		\\
55	&	2.29E+03	&		\\
65	&	3.16E+03	&	4.59E-09	\\
75	&	4.89E+03	&	1.03E-08	\\
85	&	7.87E+03	&	3.51E-08	\\
90	&	9.93E+03	&	2.58E-07	\\
95	&	1.26E+04	&	1.88E-07	\\
105	&	2.13E+04	&	1.26E-06	\\
140	&	1.24E+05	&	6.05E-04	\\
170	&	5.77E+05	&		\\
200	&	2.63E+06	&		\\
250	&	2.99E+07	&		\\
300	&	2.62E+08	&		\\
\hline					
\end{tabular}
\end{table*}

\bsp	
\label{lastpage}
\end{document}